\documentclass[a4paper,aps,prl,twocolumn,superscriptaddress,floatfix,showpacs]{revtex4}
\usepackage{amsmath}
\usepackage{graphics}
\usepackage{graphicx}
\usepackage{epsfig}
\usepackage{amsmath}
\usepackage{amsfonts}
\usepackage{amssymb}

\begin{document}

\title{First order quantum phase transitions}
\author{M. A. \surname{Continentino}}
\email{mucio@if.uff.br}
\affiliation{Instituto de F\'{\i}sica,
Universidade Federal Fluminense, Campus da Praia Vermelha,
Niter\'{o}i, 24210-340, RJ, Brazil}
\author{A. S. \surname{Ferreira}}
\affiliation{Instituto de F\'{\i}sica Gleb Wataghin, Unicamp,
Caixa Postal 6165, Campinas, SP, CEP 13083-970, Brazil}
\date{\today}

\begin{abstract}
Quantum phase transitions have been the subject of intense
investigations in the last two decades [1]. Among other problems,
these phase transitions are relevant in the study of heavy fermion
systems, high temperature superconductors and Bose-Einstein
condensates. More recently there is increasing evidence that in
many systems which are close to a quantum critical point (QCP)
different phases are in competition. In this paper we show that
the main effect of this competition is to give rise to
inhomogeneous behavior associated with quantum first order
transitions. These effects are described theoretically using an
action that takes into account the competition between different
order parameters. The method of the effective potential is used to
calculate the quantum corrections to the classical functional.
These corrections generally change the nature of the QCP and give
rise  to interesting effects even in the presence of non-critical
fluctuations.  An unexpected result is the appearance of an
inhomogeneous phase with two values of the order parameter
separated by a first order transition. Finally, we discuss the
universal behavior of systems with a weak first order zero
temperature transition in particular as the transition point is
approached from finite temperatures. The thermodynamic behavior
along this line is obtained and shown to present universal
features.
\end{abstract}


\maketitle

\section{Introduction}

Quantum phase transitions have been intensively studied in the
last two decades  \cite{sachdev}. From a pure theoretical
curiosity it became a field of intense experimental activity in
different areas of condensed matter physics. The basic concept in
this field is that of a quantum critical point (QCP). This is an
unstable fixed point which separates a phase with long range order
from a disordered phase at zero temperature \cite{Hertz,livro}. A
fundamental distinction between this type of critical point and
that associated with thermal phase transitions is the special role
that time plays as an additional dimension. This is explicitly
manifested in the quantum hyperscaling law, $2-\alpha=\nu(d+z)$
which relates the dimension of the system $d$ to the usual
exponents $\alpha$ of the singular part of the free energy density
and $\nu$ the correlation length exponent \cite{livro,mucio}. The
new feature here is the appearance of the dynamic exponent $z$
that arises from the {\em time directions}. This appears in the
suggestive  form of an effective dimension $d_{eff}=d+z$ which in
fact controls the character of the quantum fluctuations and has
most important consequences. For many problems of interest in the
laboratory $d_{eff}$ turns out to be larger or equal to the upper
critical dimension $d_c$ of the problem and consequently all
critical exponents are well known. In this case knowledge of the
dynamic exponent is sufficient to characterize the universality
class of the quantum phase transition.

The scaling form of the free energy density close to a QCP is
given by \cite{livro},
\begin{equation}
f \propto |g|^{2-\alpha}F \left[ \frac{T}{|g|^{\nu z}}, \frac{h}{
|g|^{\beta + \gamma}} \right]
\end{equation}
where {\it g} measures the distance to the QCP ({\it g }$=0$).
This expression allows to obtain the dominant thermodynamic
behavior of the system in the vicinity of the QCP. The field $h$
is that conjugated to the order parameter and the exponents
$\alpha, \beta, \gamma, \nu$ are the usual critical exponents
related by standard scaling laws \cite{livro}. Only the
hyperscaling relation is modified as discussed above. For $d_{eff}
>d_c$ there may be {\em dangerously irrelevant interactions} which
influence the critical behavior \cite{millis}, in particular along
the quantum critical trajectory $g=0, T \rightarrow 0$ and
determine the critical line of finite temperature phase
transitions \cite{millis}.

However, recent works  \cite{first} are showing that, this is not
all there is about quantum phase transitions with $d_{eff} \ge
d+z$. The region close to a QCP seems to be a turbulent zone where
many phases compete. The intensity of magnetic fluctuations near a
magnetic QCP provides an additional mechanism for pair formation
that favors the appearance of superconductivity \cite{lonz}. In
Kondo lattices local Kondo fluctuations may interfere with the
long range magnetic correlations close to the magnetic QCP
\cite{si}. In field induced Bose Einstein transitions in magnetic
systems, soft elastic modes can couple to the spin-wave
excitations with effects on the critical behavior \cite{sherman}.
Soft modes due to generic scaling invariance may also change the
critical behavior \cite{Belitz} and gauge fluctuations can affect
charged systems in the vicinity of their quantum phase transitions
\cite{haluma}. Common to all these cases is that in the region of
the phase diagram close to a QCP there is a competition of
different types of fluctuations. Our aim here is to study the
effects of this competition. In condensed matter physics, the most
well known example of fluctuations of another field interfering
with a phase transition is that of the electromagnetic field on
the thermal superconducting transition \cite{haluma}. The quantum
equivalent of this effect is known in quantum field theory as the
Coleman-Weinberg mechanism \cite{Coleman}.

\section{Coleman-Weinberg mechanism in condensed matter\label{CW-mechanism}}

In the solid-state version of the Coleman -Weinberg mechanism
\cite{Coleman}, we consider a superconductor at $T=0$ represented
by a complex field $(\varphi_{1},\varphi_{2})$ coupled to the
electromagnetic field~\cite{livro,PhysicaA}. The Lagrangian
density of the model is given by,
\begin{eqnarray}
&&L
=-\frac{1}{4}(F_{\mu\nu})^{2}\!+\!\frac{1}{2}(\partial_{\mu}\varphi
_{1}\!+\!qA_{\mu}\varphi_{2})^{2} \nonumber\label{SClagrangian}\\
&&  +\frac{1}{2}(\partial_{\mu}\varphi_{2}-qA_{\mu}\varphi_{1})^{2} \nonumber\\
&& - \frac{1}{2}m^{2}(\varphi_{1}^{2}+\varphi_{2}^{2})-\frac{\lambda}%
{4!}(\varphi_{1}^{2}+\varphi_{2}^{2})^{2}.
\end{eqnarray}
We are using $\hbar=c=1$ units and the indices $\mu,\nu$ run from
$0$ to $d=3$. In Eq.~(\ref{SClagrangian}) space and time are
isotropic and consequently the dynamic critical exponent $z=1$.
For a neutral superfluid $(q=0)$ the system decouples from the
electromagnetic field and has a continuous, zero temperature
superfluid-insulator transition at $m^{2}=0$ (see
Fig.~\ref{fig1}).

The method of the effective potential~\cite{livro} yields the
quantum corrections to the action given by the Lagrangian density
of Eq.~(\ref{SClagrangian}). At $T=0$ in the one loop
approximation, the effective potential close to the transition
($m\approx0$) is given by~\cite{livro}
\begin{equation}
V_{eff}\!=\!\frac{1}{2}m^{2}\varphi^{2}\!-\!\frac{m^{2}}{4\langle
\varphi\rangle^{2}}\varphi^{4}\!+\!\frac{3q^{4}}{64\pi^{2}}\varphi^{4}\left[
\ln\left(  \frac{\varphi^{2}}{\langle\varphi\rangle^{2}}\right)
\!-\!\frac {1}{2}\right]  \label{eff2}
\end{equation}
where $\langle\varphi\rangle$ is the minimum of the effective
potential.  We can show \cite{jstat,S.Stat.Comm.} that the
condition for such a minimum to exist is $\lambda_{L}\ll\xi$,
where $\xi$ is the coherence length and $\lambda_{L}$ the London
penetration depth as usually defined for Ginzburg-Landau models.
In this case we find \cite{jstat} that at a critical value of the
mass, given by
\begin{equation}
\label{shift}
m_{c}^{2}=\frac{3q^{4}}{32\pi^{2}}\langle\varphi\rangle^{2}%
\end{equation}
there is a first order transition to a superconducting state.
Notice that the transition in the neutral superfluid ($q=0$) is
continuous rather than first order and takes place at $m^{2}=0$.
Therefore,  the coupling to the electromagnetic field in the
charged superfluid lead to symmetry breaking, shifting the
transition of the neutral superfluid (see Fig.~\ref{fig1}) and
changed its nature from continuous to first order.  The shift of
the transition Eq. (\ref{shift}) depends on the coupling of the
order parameter to the soft modes, in the present case, the charge
$q$ of the Cooper pairs.

Finally, we point out that the coupling $\lambda$ has disappeared
from Eq. (\ref{eff2}) for the effective potential due to {\em
dimensional transmutation} \cite{Coleman}.

\section{Competition between superconductivity and antiferromagnetism}

In this section we show that weak first order quantum phase
transitions  (WFOQPT) and spontaneous symmetry breaking can also
occur due to the competition between different types of
instabilities in the same region of the phase diagram.

We consider a Ginzburg-Landau model which is appropriate to
describe the competition between superconductivity (SC) and
antiferromagnetism in a heavy fermion metal. The model contains
three real fields. Two fields, $\phi_{1}$ and $\phi_{2}$,
correspond to the two components of the superconductor order
parameter. The other field $\phi_{3}$, for simplicity represents a
one component antiferromagnetic (AF) order parameter. The free
functional of the magnetic part~\cite{PRB} takes into account the
dissipative nature of the paramagnons near the magnetic phase
transition in the metal~\cite{Hertz} and is associated with the
propagator,
\begin{equation}
D_{0}(\omega, {\bf q})=\frac{i}{i|\omega|\tau -q^{2}-m_{p}^{2}}
\label{propagatorAF}
\end{equation}
where $\tau$ is a characteristic relaxation time and $m_{p}^{2}$
gives the distance to the magnetic transition. The quadratic form
of the superconductor, the same used in the previous section,  is
given by,
\begin{equation}
G_0(k)=G_0(\omega,{\bf q})=\frac{i}{k^2-m^2}.
\end{equation}
The part of the action associated with the classical potential is,
\begin{eqnarray}
\label{pot}V_{cl}(\phi_{1},\phi_{2},\phi_{3})=\frac{1}{2}m^{2}(\phi_{1}
^{2}+\phi_{2}^{2})+\frac{1}{2}m_{p}^{2}\phi_{3}^{2} \nonumber\\
+ V_{s}(\phi_{1},\phi_{2}) + V_{p}(\phi_{3}) +
V_{i}(\phi_{1},\phi_{2} ,\phi_{3}),
\end{eqnarray}
where the self-interaction of the superconductor field is
\begin{equation}
V_{s}(\phi_{1},\phi_{2})=\frac{\lambda}{4!}(\phi_{1}^{2} +
\phi_{2}^{2})^{2},
\end{equation}
and that of the antiferromagnet,
\begin{equation}
V_{p}(\phi_{3})=\frac{g}{4!}\phi_{3}^{4}.
\end{equation}
Finally, the last term is the (minimum) interaction between the
relevant fields,
\begin{equation}
\label{coupling}V_{i}(\phi_{1},\phi_{2},\phi_{3})=u(\phi_{1}^{2}+\phi_{2}%
^{2})\phi_{3}^{2}.
\end{equation}
This term is the first allowed by symmetry on a series expansion
of the interaction. Notice that for $u>0$, which is the case here,
superconductivity and antiferromagnetism are in competition and
this term does not break any symmetry of the original model.
However, including quantum fluctuations we show that spontaneous
symmetry breaking can occur in the normal phase separating the SC
and AF phases.

\begin{figure}[tbh]
\begin{center}
\includegraphics[width=8cm,height=1.5cm]{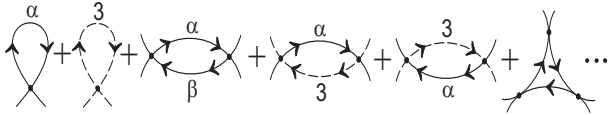}
\end{center}
\caption{One loop diagrams contributing to the effective
potential. } \label{fig loops}
\end{figure}

The first quantum correction to the potential can be obtained by
the summation of all one loop diagrams~(Fig.~\ref{fig loops}).

We apply the general method proposed by Coleman~\cite{Coleman}
with minimum modifications to account for the different nature of
the propagators in our problem. The sum over the field indices can
be easily done  if we define a vertex matrix $\mathbf{M}$, given
by
\begin{equation}
\label{M}[M]_{lm}=-iK^{l}_{0}\frac{\partial^{2}V_{cl}}{\partial\phi
_{l}\partial\phi_{m}}\Big|_{\{\phi\}=\{\phi_{c}\}}
\end{equation}
and then take the trace. In Eq.~(\ref{M}) the propagators
(${K^{l}_{0}=G_{0} \mbox{ or } D_{0}}$) are incorporated in the
definition of the matrix. We draw the loops with arrows and choose
the outgoing propagator of each vertex to be included in the
associated element. The matrix $\mathbf{M}$ is then obtained
deriving the classical potential with respect to the fields
$\{\phi\}$ and taking the values of these derivatives at the
classical values of the fields, $\{\phi_{ic}\}$. The sum of
diagrams with the correct Wick factors is formally done in
momentum space and using the property of the trace
\begin{equation}
\mbox{Tr}[\ln(1-M)]=\ln\mbox{det}[(1-M)],
\end{equation}
we get
\begin{equation}
V^{(1)}[\phi_{c}]=\frac{i}{2}\hbar\int d^{4} k \ln\det\left[
1-M(k)\right].
\end{equation}
The $3\times3$ matrix $\mathbf{M}$ can be simplified if we choose
the classical minimum of the superconductor fields imposing
${\phi_{2c}=0}$ (this can be done because the minimum depends only
on the modulus $\phi_{1c}^{2}+\phi_{2c}^{2}$). Hence, rotating to
Euclidean space, so that, $k^{2}=\omega^{2}+q^{2}$ and using
$\hbar=1$ units the first quantum correction can be written as
\begin{eqnarray}
&&V^{(1)}(\phi_{1c},\phi_{3c})=\frac{1}{2}\int\frac{d^4k}{(2\pi)^4}\left\{
\ln\left(1+\frac{A(\phi_{1c},\phi_{3c})}{k^2+m^2}\right)
\right.   \nonumber \\
&& +
\left.\ln\left[\left(1+\frac{B(\phi_{1c},\phi_{3c})}{k^2+m^2}\right)
\left(1+\frac{C(\phi_{1c},\phi_{3c})}{|\omega|\tau+q^2+m_p^2}\right)
\right.\right. \nonumber \\
&& \left.\left.
-\left(\frac{D^2(\phi_{1c},\phi_{3c})}{(k^2+m^2)(|\omega|\tau+q^2+m_p^2)}\right)
\right]\right\} \label{QC}
\end{eqnarray}
where
\newpage
\begin{eqnarray}
A(\phi_{1c},\phi_{3c})=(\lambda/6)\phi_{1c}^{2}+2u\phi_{3c}^{2}\\
\nonumber \\
B(\phi_{1c},\phi_{3c})=(\lambda/2)\phi_{1c}^{2}+2u\phi_{3c}^{2}\\
\nonumber \\
C(\phi_{1c},\phi_{3c})=2u\phi_{1c}^{2}+(g/2)\phi_{3c}^{2}\\
\nonumber \\
D(\phi_{1c},\phi_{3c})=4u\phi_{1c}\phi_{3c}\label{D coef.}%
\end{eqnarray}
The total effective potential with first order quantum corrections
is then given by
\begin{equation}
V_{eff}(\phi_{1c},\phi_{3c})=V_{cl}(\phi_{1c},\phi_{3c})+V^{(1)}(\phi_{1c}
,\phi_{3c})
\end{equation}
where $V_{cl}$ is the classical potential of Eq.~(\ref{pot}) and
$V^{(1)}$ is the first quantum correction of order $\hbar$ of
Eq.~(\ref{QC}).

\subsection{Superconducting transition\label{SCtransition}}

We first consider the effect on the superconductor transition in
HF in the presence of antiferromagnetic paramagnons $(\phi_{1c}
\ne 0, \phi_{3c}=0)$. Detailed calculation of the effective
potential have already been presented~\cite{PRB}. The general
result is given by,
\begin{equation}
\label{potSC}V_{eff}(\phi_{c})
\approx\frac{1}{2}M^{2}\phi_{1c}^{2}+a
m_{p}^{2}\phi_{1c}^{2}|\phi_{1c}|+\frac{\tilde{\lambda}}{4!}\phi_{1c}^{4}+\mathcal{O}(\phi
^{5}).
\end{equation}
In Eq.~(\ref{potSC}), $M$ is a renormalized superconducting mass
and $\tilde{\lambda}$ a renormalized coupling, of the same order
of the bare coupling $\lambda$. The new coupling $a$ introduced by
fluctuations can produce a symmetry breaking in the normal state
extending the SC region in the phase diagram at $T=0$. The same
mechanism turns the superconducting transition, which was
continuous before coupling to the paramagnons, to weak first order
with a small latent heat\cite{S.Stat.Comm.}. Introducing again the
coherence length $\xi$  and the London penetration depth
$\lambda_{L}$ we can show that the condition for the existence of
minima away from the origin is equivalent to $\lambda_{L}\ll\xi$
as in the previous case~\cite{PRB}. It's also interesting to
notice that the first order transition is produced by the cubic
term in Eq.~(\ref{potSC}) and this term is proportional to the
{\it magnetic mass} $m_{p}$. If magnetic fluctuations were
critical, i.e.,  $m_{p}=0$, the only effect of the coupling would
appear as a term proportional to $\phi^{5}$. This term is usually
neglected since its power is higher than those initially
considered in the classical potential and usually insufficient to
create new minima around the origin. Therefore, if AF fluctuations
were critical the effects of the quantum corrections in the
transition could be neglected.

\subsection{Antiferromagnetic transition}

We now study the effect of SC fluctuations in the magnetic
transition $(\phi_{1c} = 0, \phi_{3c}\ne 0)$. We consider two
kinds of quadratic forms associated with the free superconducting
fields. The first is the usual Lorentz invariant free action used
before. Next, we work with another free action which takes into
account dissipation and is associated with a $z=2$ dynamics
\cite{Ramazashvili,Kirkpatrick,Sigrist}.

Let's first study the case of the Lorentz invariant propagator.
Close to the AF transition we have~\cite{PRB}
\begin{eqnarray}
\label{potAF} &&V_{eff}(\phi_{3c})
\approx\frac{1}{2}M_{p}^{2}\phi_{3c}^{2}
+\frac{\tilde{g}}{4!}\phi_{3c}^{4}+\tilde{u}^{2}\phi_{3c}^{4}\ln\left(
\frac{\phi_{3c}^{2}}{\langle\phi_{3}\rangle^{2}}\right) \nonumber\\
&&+  \tilde{u}m^{2}\phi_{3c}^{2}\ln{\left(
\frac{\phi_{3c}^{2}}{\langle \phi_{3}\rangle^{2}}\right)  }
\end{eqnarray}
Notice  from Eq.~(\ref{potAF}) that if the SC fluctuations were
critical, i.e. $m^{2}=0$, we would obtain the same result of
Eq.~(\ref{eff2}), i.e., a fluctuation induced quantum first order
transition. However, if the SC fluctuations are not critical but
close to criticality the last term of Eq.~(\ref{potAF}) may become
important. Of course, its relevance depends on the strength of the
renormalized coupling $\tilde{u}$ and the results considering this
term lead to new and interesting changes in the ground state. We
obtain, besides the two finite minima of the Coleman-Weinberg
potential of Eq.~(\ref{eff2}),  two extra minima very close to the
origin~\cite{PRB}. The  states associated with these minima have a
small value of the order parameter, the sub-lattice magnetization.
An additional first order transition occurs when the other two
minima away from the origin become the stable ones as the system
moves away from the superconductor instability. This transition is
from a small moment AF (SMAF) to a large moment AF (LMAF) and
occurs even before the continuous mean field transition. When we
move away from the magnetic transition,  i.e., towards the
superconducting instability,  the strength of this new term
decreases with the value of $m^{2}$ and the two new minima move to
the origin producing a normal state with vanishing sub-lattice
magnetization again. Notice that the SMAF phase is obtained
because the magnetic order parameter couples to superconducting
fluctuations which are non-critical. Critical fluctuations yield
the same results of section~\ref{CW-mechanism}.

\begin{figure}[h]
\begin{center}
\includegraphics[height=5.5cm]{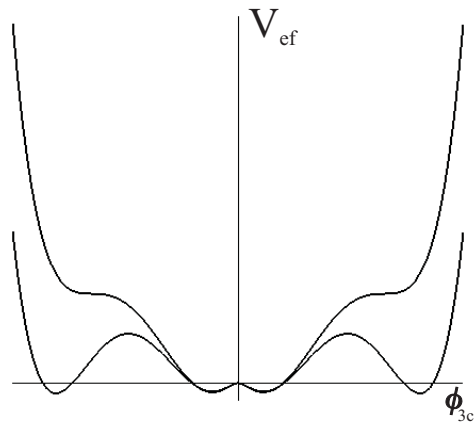}
\end{center}
\caption{ New minima appear in the potential for $um^2 \ne 0$. The
effective potential is shown here for two situations: for
$M_p^2=M_p^{c2}$, where the first order transition from LMAF to
SMAF occurs and these two states become degenerate and for the
spinodal point at which the LMAF becomes unstable inside the SMAF
phase. }%
\label{diagram}
\end{figure}

Now, for many cases of interest, SC fluctuations are better
described by a dissipative propagator associated with a $z=2$
dynamics \cite{Ramazashvili,Kirkpatrick} similar to
Eq.~(\ref{propagatorAF}). This is useful to account for pair
breaking interactions, as magnetic impurities that can destroy
superconductivity~\cite{Sigrist}. It is given by,
\begin{equation}
G_{0}(\omega, {\bf
q})=\frac{i}{i|\omega|\tau^{\prime}-q^{2}-m^{2}}.
\end{equation}
The parameter $m^{2}$ is still related to the  distance from the
SC phase transition and we have a relaxation time $\tau^{\prime}$.
Calculation of the effective potential is very similar to the
previous cases and the result has the form
\begin{equation} \label{Veffz=2}
V_{eff}=\frac{1}{2}M_{p}^{2}\phi_3^{2}+\frac{1}{4!}\tilde{g}\phi_3^{4}+\frac
{1}{15\pi^{2}}(2u\phi_3^{2}+m^{2})^{5/2}
\end{equation}
with a renormalized {\em magnetic mass} $M_{p}$ and  coupling
$\tilde{g}$. Quantum corrections can once again produce a weak
first order transition. An analysis of the extrema of the
effective potential Eq.~(\ref{Veffz=2}) shows that the transition
can be first order depending on the coupling values. The
appearance of SMAF phases is not possible in this case.

\section{Coupling to local modes}

We now study the coupling of antiferromagnetic (AF) fluctuations
to local modes in a three dimensional system. This model is useful
to describe the QCP of heavy fermions where local modes can
coexist with  antiferromagnetic fluctuations. The local modes can
be either Kondo  \cite{si} or valence fluctuations \cite{miyako}.
The local propagator in Euclidean space is written as
\begin{equation}
G_L(\omega)=\frac{1}{m_L^2+|\omega|/\tau}
\end{equation}
where $ m_L^2$ gives the distance to the QCP and $\tau$ is
associated with the lifetime of the excitations. For the AF
paramagnons, we have
\begin{equation}
G_p(\omega,q)=G_p(k)=\frac{1}{m_p^2+q^2+|\omega|}
\end{equation}
The action  is
\begin{eqnarray}
S&=&\int d^4k [ G_p^{-1}(k)|\psi|^2 +\frac{g}{12}|\psi|^4
+G_L^{-1}(\omega)|\phi|^2 \nonumber \\
&&+\frac{\lambda}{12}|\phi|^4+ \frac{u}{2}|\phi|^2|\psi|^2 ].
\end{eqnarray}
Let us consider one component fields, the classical potential (not
including the mass terms) is given by,
\begin{equation}
V_{cl}(\phi,\psi)=\frac{g}{12}\psi^4
+\frac{\lambda}{12}\phi^4+\frac{u}{2}\phi^2\psi^2
\end{equation}
and the effective potential by,
\begin{eqnarray}
 &&V^{(1)}(\psi_{c})=\frac{1}{2}\int\frac{d^4k}{(2\pi)^4}
 \ln\left(1+\frac{u\psi_c^2}{m_L^2+|\omega|/\tau}\right)
 \nonumber \\
&&+\frac{1}{2}\int\frac{d^4k}{(2\pi)^4}
 \left(1+\frac{g\psi_c^2}{m_p^2+q^2+|\omega|}\right)
 + \cdots \label{V1simp}
\end{eqnarray}
where the dots represent counter-terms \cite{futuro}. The
integration in $\omega$ in the first integral can be performed
using a cut-off, $\Lambda^z-m_L^2{\tau}$ ($z=2$). However, the
integral over $d{\bf q}$ diverges when $\Lambda \rightarrow
\infty$ and this term is non-renormalizable unless we introduce a
cut-off $q_c$ and for later purposes $\tilde\tau=(4q_c^3)/(3\pi)$.

\subsection{Case $m_L^2 \gg \tilde\tau u^2$}

In this case the analysis of the renormalized effective potential
shows that, as for the classical potential, there is a second
order phase transition at $m_p^2=0$ and the coupling to the local
modes has no significant effects.

\subsection{Case $m_L^2 \ll \tilde\tau u^2$}

\begin{figure}[h]
\begin{center}
\includegraphics[height=5.5cm]{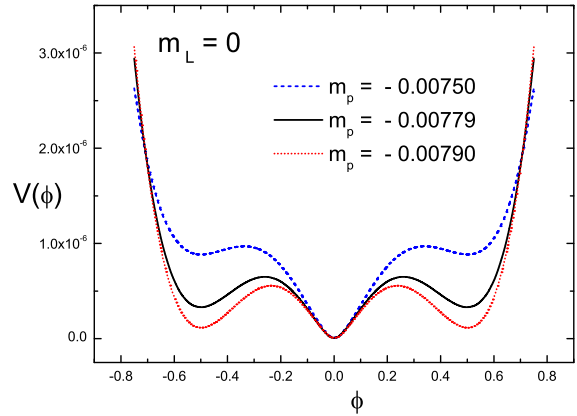}
\end{center}
\caption{ For $m_L^2 = 0$ there is a first order transition to the
antiferromagnetic phase~\cite{futuro}.  }%
\label{mlzero}
\end{figure}

\begin{figure}[h]
\begin{center}
\includegraphics[height=5.5cm]{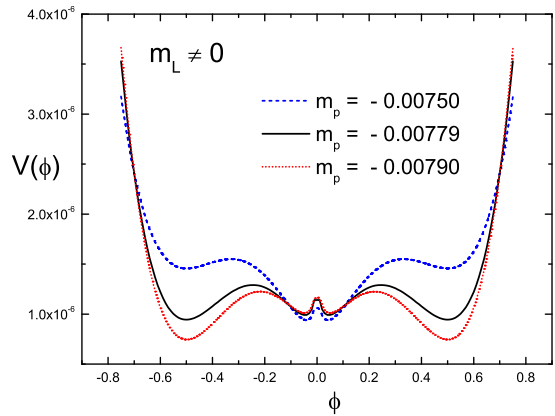}
\end{center}
\caption{ For $m_L^2 \ne 0$ new minima appear close to the origin~\cite{futuro}. }%
\label{two}
\end{figure}

This is the most interesting case. The renormalized effective
potential after dimensional transmutation is given by,
\begin{eqnarray}
&&V_{eff}(\psi)=\tilde{M}_p^2 \psi^2 +\tilde{G}\psi^4
+\frac{\pi^2}{(2\pi)^4}\left[-\tilde\tau
u\psi^2\ln\left(\frac{\psi^2}{\langle \psi\rangle^2}\right)
\right.  \nonumber \\
&&\left. +\frac{8}{15}\left(-\frac{6m_p^2 \pi^2 +
\tilde{\tau}u}{\langle \psi \rangle^2 \pi^2}
\psi^2+m_p^2\right)^{5/2}\!\!\!\!\!\!- \tilde\tau m_L^2\ln
(u\psi^2) \right]
\end{eqnarray}
where the renormalized quantities $\tilde{M}_p^2$ and $\tilde{G}$
are independent of $g$~\cite{futuro}. In this case we find there
is no transition for $m_p^2 \ge 0$, such that, there is no dynamic
symmetry breaking. A first order transition occurs for negative
values of $m_p^2$, at a  critical value ${m_p^c}^2$ which has to
be obtained numerically. The limit of stability of the minima of
the effective potential (spinodal) is given by,
$m^2_{p,sp}=-(\tilde\tau u)/(4\pi^2)$, with $m^2_{p,sp} >
{m_p^c}^2$. For the particular case $m_L^2 =0$ there is a first
order transition between the paramagnetic and antiferromagnetic
phases, as shown in Fig. \ref{mlzero}. Then, even in $3d$, when
both transitions coincide, the magnetic transition becomes first
order due to the effect of the critical local fluctuations.
Now, for $m_L^2$ small but $\ne 0$, the phase diagram changes
drastically since new minima appear in the potential close to the
origin. These minima which correspond to small values of the order
parameter (SMAF) can coexist with those associated with the large
values of this parameter (LMAF). As $m_p^2$ further decreases
there is a first order transition between the SMAF and LMAF phases
(see fig. \ref{two}).

\section{Scaling at a weak first order quantum transition}

At a first order transition there is no true critical behavior
since the correlation length does not diverge. However it turned
out to be useful to develop a scaling approach for these
transitions in the classical case \cite{berker}. As we show below
the same is true for first order quantum phase transitions. This
is particularly useful for WFOQPT where we expect the correlation
length and characteristic time to become very large. The best way
to introduce these ideas is to consider a specific case, for
example, the Coleman-Weinberg transition of section
\ref{CW-mechanism}.

Introducing a parameter $g=m^{2}-m_{c}^{2}$ which measures the
distance to the first order transition at $m_c$, we can write
Eq.~\ref{eff2} for the effective potential at $T=0$ and for $g$
small as,
\begin{equation}
V_{eff}= \frac{1}{4} \langle\varphi\rangle^{2}|m^{2}-m_{c}^{2}| \propto |g|^{2-\alpha}%
\end{equation}
with the exponent $\alpha=1$ reflecting the fact the transition is
first order~\cite{PhysicaA}. The associated \emph{latent heat} is
$L_h =\frac{1}{4} m_c^2\langle\varphi\rangle^{2}$. Spinodal points
at $T=0$ can also be calculated~\cite{PhysicaA}.

\begin{figure}[h]
\begin{center}
\includegraphics[height=5.5cm]{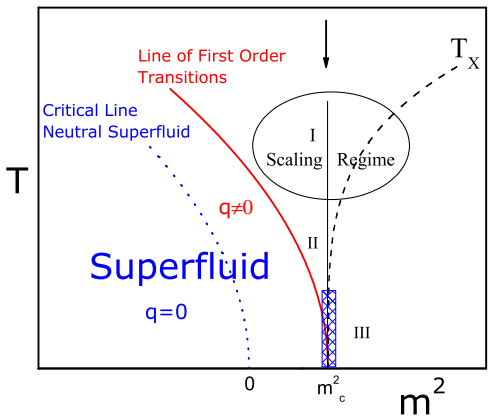}
\end{center}
\caption{Phase diagram of a charged superfluid coupled to photons.
For completeness we show also the critical line of the neutral
superfluid. Along the trajectory $m^{2}=m_{c}^{2}$ one can
distinguish different regimes as
explained in the text.}%
\label{fig1}%
\end{figure}

The finite temperature case can be studied replacing the frequency
integrations in the calculation of the effective potential by sums
over Matsubara frequencies~\cite{livro}. The effective potential
at finite temperatures close to the transition can be written
as~\cite{PhysicaA}
\[
V_{eff}(T)=\frac{1}{4}m^{2}\langle\varphi\rangle^{2} |g| \left[
1\!+\!\frac{2}{\pi^{2}m^{2}\langle\varphi\rangle^{2}}\frac{T^{d+1}}
{|g|}I_{d}\!\left(  \frac{M(\varphi)}{T}\right)  \right]
\]
where the integral $I$ is given by
\begin{equation} \label{I}
I_{d}(y)=\int_{0}^{\infty}dxx^{d-1}\ln{\left[
1-e^{-\sqrt{x^{2}+y^{2}} }\right]  }.
\end{equation}
and $M(\varphi)=m^{2}+q^{2}\varphi^{2}$. The function
$I_{3}(y)=I(y)$ can be obtained numerically integrating
Eq.~(\ref{I}).

The finite temperature phase diagram is shown in Fig.~\ref{fig1}.
For completeness we show the critical line of the neutral
superfluid,
$T_{SF} \propto |m^{2}|^{\psi}$, which is governed by the shift exponent ${\psi}%
^{-1}=z/(d+z-2)=1/2$ in $d=3$ (see Ref.~\cite{livro}). The new
line of first order transitions is given by $T_{c} \propto
\sqrt{m_{c}^{2}-|m^{2}|}$.

We will now consider the system \emph{sitting} at the new quantum
phase transition point, i.e., at $m^{2}=m_{c}^{2}$ and decrease
the temperature. For high temperatures, $T \gg m_{c}$, which
corresponds to the regime I of Fig.~\ref{fig1}, the function
$I_{3}(y)$ saturates, $I_{3}(y < 0.12 )\approx- 2.16$. In this
case the effective potential,
\begin{eqnarray*}
V_{eff}(T)
\approx\frac{1}{4}m^{2}\langle\varphi\rangle^{2}|g|\bigg\{ 1-\frac
{4.32}{\pi^{2}m^{2}\langle\varphi\rangle^{2}}\frac{T^{d+1}}{|g|}\bigg\}
\end{eqnarray*}
and can be cast in the scaling form,
\[
V_{eff}(T) \propto|g|^{2-\alpha} F \left[  \frac{T}{T_{\times}}
\right]  .
\]
with $F(0)=$ $constant$. This scaling form is reminiscent of that
for the free energy close to a quantum critical point. In the
present case of a discontinuous zero temperature transition, the
critical exponent~\cite{PhysicaA} $\alpha=1$ and the
characteristic temperature is,
\[
T_{\times} \propto|g|^{\nu z} \propto|g|^{\frac{z}{d+z}} =
|g|^{\frac{1}{d+1}}=|g|^{\frac{1}{4}}
\]
with $\nu=1/(d+z)$~\cite{PhysicaA}. In this regime I or scaling
regime, along the line $m^{2}=m_{c}^{2}$ shown in Fig.~\ref{fig1},
the free energy density has therefore the scaling form $f(m=m_{c},
T) \propto T^{(d+z)/z} $ and the specific heat is given by,
\begin{equation}
C/T\big|_{(m=m_{c}, T)} \propto T^{\frac{d-z}{z}}.
\end{equation}
Then the thermodynamic behavior along the line $m^{2}=m_{c}^{2}$
in regime I ($T \gg m_{c}$) is \emph{the same} as when approaching
the quantum critical point of the neutral superfluid, along the
critical trajectory $m^{2}=0$. The system is unaware of the change
in the nature of the zero temperature transition and at such high
temperatures charge is irrelevant.

When further decreasing temperature along the line
$m^{2}=m_{c}^{2}$ there is an intermediate, non-universal regime
(regime II in Figs.~\ref{fig1}). In the present case for $T \sim
m_{c} $ the specific heat $C/T^{d/z} \propto\ln
T$~\cite{PhysicaA}.

Finally, at very low temperatures, for $T << m_{c}$ and
$m^{2}=m_{c}^{2}$, i.e., in regime III of Fig.~\ref{fig1}, the
specific heat vanishes exponentially with temperature, $C/T^{d/z}
\propto\exp(-m_{c}/T)$. The gap for thermal excitations is given
by the shift $m_{c}$ of the quantum phase transition. The
correlation length which grows along the line $m^{2}=m_{c}^{2}$
with decreasing temperature reaches saturation in regime III at a
value $\xi_{S}$ which is function of the inverse of the gap. The
exponential dependence of the specific heat is due to gapped
excitations inside superconducting bubbles of finite size $\xi_{S}
\propto m_c^{-1/z}$.

Although the results above have been obtained for the model of
section~\ref{CW-mechanism}, the behavior in the scaling regime I
and III should be universal and characteristic of any weak first
order quantum transition. Notice that the relevant critical
exponents which determine the scaling behavior in particular in
regime I are those associated with the QCP of the {\em uncoupled
system} which in the present case is the neutral superfluid.

\acknowledgements{MAC would like to thank the Brazilian agencies
CNPq and FAPERJ for partial financial support. ASF acknowledges
the Brazilian agency FAPESP for financial support.}

\end{document}